\documentclass[aps,preprint]{revtex4}%
\usepackage{amsfonts}
\usepackage{amsmath}
\usepackage{amssymb}
\usepackage{graphicx}%
\begin{document}
\preprint{05RBP03/Cal02 (sup100calc2.tex)}
\title[(100) chalcopyrite surface states]{(100) ideal-surface band structure for the series of Cu-based $A^{I}%
B^{III}C_{2}^{VI}$ \ chalcopyrites. }
\author{H. Tototzintle-Huitle$^{\ast}$ and R. Baquero}
\affiliation{Departamento de F\'{\i}sica, Cinvestav. A.Postal 14-780, 07000 M\'{e}xico D.F.}

\begin{abstract}
We use the Surface Green%
%TCIMACRO{\U{b4}}%
%BeginExpansion
\'{}%
%EndExpansion
s Function Matching (SGFM) method and a tight-binding hamiltonian to calculate
the (100)-surface electronic band structure and local density of states of the
series of Cu-based $A^{I}B^{III}C_{2}^{VI}$ chalcopyrites . We find four
surface states in the optical gap energy region of s-p character and three
surface states in the conduction band region of p-character. We show the
trends of different characteristics within the series by means of figures and
tables so that the quantitave behavior can be evaluated as well. We did not
find Frontier Induced Semi-Infinite states of non-dispersive character in the
studied range of energy within the valence band as we found in the case of the
(112) surface electronic band structure for $CuInSe_{2}$.

* also at Convestav, Quer\'{e}taro.

corresponding author: rbaquero@fis.cinvestav.mx

Shell document for REV\TeX{} 4.

\end{abstract}
\volumeyear{year}
\volumenumber{number}
\issuenumber{number}
\eid{sup100calc2.tex}
\date[Date text]{15 may 2005}
\received[Received text]{date}

\revised[Revised text]{date}

\accepted[Accepted text]{date}

\published[Published text]{date}

\startpage{1}
\endpage{ }
\maketitle

\section{Introduction}

Semiconducting chalcopyrites are interesting for their non-linear optical
properties and for their possible technological applications \cite{ref1,
optprop}. Chalcopyrites are used in processes as chemisorption, catalysis and
bioleaching \cite{ref2}, solar cells \cite{solarcell} and other applications
\cite{ref4,diode}. Surfaces, interfaces, quantum wells and superlattices with
chalcopyrites as one of the elements are of interest in this context.

At the basis of a proper description of these systems is an accurate enough
calculation of the electronic band structure. The first calculation of the
electronic structure for the series of ternary Cu-based chalcopyrites was
presented by Jaffe and Zunger \cite{zungergeneric}. It was calculated
self-consistently within the density functional formalism. They present a
generic diagram of the bulk electronic band structure for the Cu-based
chalcopyrites. The electronic band structure of bulk CuInSe$_{2}$ and of its
(112) surface has been studied later on by Rodr\'{\i}guez et al. \cite{prbJA}
using the tight-binding method. They implemented a set of tight-binding
parameters for the series of chalcopyrites $CuB^{III}C^{VI}$ with $B=Al,Ga,In$
and $C=S,Se,Te$ that describe quite accurately the bulk electronic band
structure and have been used successfully \cite{tesisJA, rmfJA} as input to
calculate the surface electronic band structure using the Surface Green's
Function Matching (SGFM) method \cite{sgfm}. Interfaces \cite{baqinterf} and
superlattices \cite{baqsuperlat} can also been described successfully by this method.

\textit{Ab initio} methods have been used to calculate the bulk electronic
band structure although sometimes the semiconductor optical gap is not
reproduced accurately. Recently these methods have been used successfully to
study ferromagnetism of Mn-substituted Cu-based chalcopyrites \cite{zunger}.
The main interest of these works is related to spintronics. Mn-doped
II-IV-V$_{2}$ chalcopyrites have been studied in ref. \cite{tailoring}.
Comparison of predicted ferromagnetic tendencies of Mn substituting the Ga
site in II-V's and I-III-VI$_{2}$ chalcopyrites semiconductors is done in
\cite{tendencies} and the site preference for Mn substitution in spintronic
chalcopyrite semiconductors is explored in \cite{sitepreference}. Other recent
studies on chalcopyrites are on magnetic defects in a photovoltaic material
(Fe-doped \textit{CuInSe}$_{2})$ \cite{photovoltaic} and the effect of
hydrogen (H) on Cu-based chalcopyrites. Actually, for \textit{CuGaSe}$_{2}%
$\textit{ }it is found that H is a deep donor while for \textit{CuInSe}$_{2}$
it is a shallow one. The conductivity can be changed from p-type to n-type
either by introducing hydrogen to the sample as a dopant or by passivating the
defects. The final result depends on details of the minimum of the conduction
band of the chalcopyrite and on stoichiometry \cite{hydrogen}. \textit{Ab
initio} methods have also been used to study other bulk properties as
intrinsic point defects in \textit{CuInSe}$_{2}$ \cite{pointdefects}, anion
vacancies as a source of persistent photoconductivity in \textit{CuGaSe}$_{2}$
and \textit{CuInSe}$_{2}$ \cite{vacancies}, and to establish the existence of
an energetic barrier for holes arriving from the grain interior to the grain
boundaries in \textit{CuInSe}$_{2}$ \cite{grainboundary}.

There are in the literature several other studies on chalcopyrites of the type
$A^{I}B^{III}C^{VI}$. The stability of this crystal phase\ has been studied
and its possible coexistence with a phase of CuAu-type has been predicted in
some cases \cite{ref7}. Structural and dynamic properties of the lattice of
some chalcopyrites using the perturbation method to the density functional are
presented in reference \cite{ref10}. They calculate the static and high
frequency dielectric tensor and get a good agreement with infrared and Raman
experimental results. Order-disorder transitions can be induced by doping the
sample with a magnetic ion as Mn to a dilute magnetic semiconductor with a
slightly modified crystal structure \cite{ref8}.

Reference \cite{ref11} deals with a systematic treatment of the stability as a
function of the concentration of the components in $ZnGeAs_{2}$ / $GaAs$ in
the (001) direction. The effect of anion and tetragonal distortions on the
(112) surface has been recently studied for a series of Cu-based chalcopyrites
by Tototzintle \textit{et al}. \cite{ref18}. For $CuInSe_{2}$, the electronic
states of the ideal surfaces (100) and (112) has been calculated by
Rodr\'{\i}guez\textit{ et al. }\cite{brazilian}\textit{, }the reconstruction
of the polar (112) and (-1-1-2) surfaces versus the non-polar (220) one has
been studied by Zhang and Wei \cite{wei}. Also the defect induced non-polar to
polar transition at the (112) surface is considered in reference \cite{ref17}.
On the basis of atomic chemical potentials, they show that the (112) surface
is the more stable. No extensive calculation on the (100) surface for the
series we deal with in this work is known to us. The mechanism of linear and
non-linear optical effects of AgGaX$_{2}$ ($X=S,Se,Te$) has been recently
studied by Bai \textit{et al. }\cite{linear}.

Recent experiments on Cu-based chalcopyrites include the study of the
influence of hydrogen \cite{hydrogenexp}, Raman \cite{raman}, atomic defects
and non-stoichiometry \cite{defectsexp} and muon charge states \cite{muones}
all in $CuInSe_{2}$; quasi real-time Raman studies on the growth of Cu-In-S
thin films \cite{realtime} and temperature dependent Hall measurements of
intrinsic defects in CuGaSe$_{2}$ \cite{hall}.

In this paper we use the tight-binding hamiltonian and the parameters
previously constructed \cite{tesisJA, prbJA, rmfJA} and the Surface Green's
function matching method (SGFM) \cite{sgfm} to study the electronic band
structure of the ideal (001) surface for the series of the Cu-based
chalcopyrites $CuB^{III}C^{VI}$ where $B=Al,Ga,In$; $C=S,Se,Te$. The method
has been proved accurate enough to describe other systems as II-VI
\cite{ref15, angela} and III-V zincblende \cite{decoss} and wurzites
\cite{ref16} semiconductors. The Surface Green's function matching method
\cite{sgfm} starts with the bulk tight-binding parameters as input and takes
care by itself for the proper matching condition that describes the surface.

The tight-binding method \cite{slaterkoster}, the construction of the
hamiltonian and the Surface Green's function matching method are described
extensively enough elsewhere \cite{sgfm, alejandro, quintanar, prbJA, rmfJA,
ref18}. We therefore limit ourselves in the rest of the paper to our new results.

\section{The (100) bulk projected bands for the series $CuB^{III}C^{VI}$ where
$B=Al,Ga,In$; $C=S,Se,Te$.}

Previous to the electronic density of states for the (100) ideal surface, it
is convenient to describe the general trends in the (100) bulk-projected
electronic band structure for the series. A generic diagram emerges from the
bulk electronic band structure \cite{baquerobulk} similar to the one
elaborated previously by Jaffe and Zunger \cite{zungergeneric}. As we already
mentioned, we used the SGFM method \cite{sgfm} to calculate the (100)-surface
Green%
%TCIMACRO{\U{b4}}%
%BeginExpansion
\'{}%
%EndExpansion
s function and from the poles of its real part we calculated the band
structure that we present on Figs.1-3.. The shaded areas are the bulk bands.
The origin is at the top of the valence band. Notice that we have drawn the
bands from -6 eV to +10 meV. The bulk bands presented here as shaded areas are
obtained from the (100)-electronic band structure projected into the bulk
(fifth layer). The bands are drawn through the high-symmetry points
$X-\Gamma-$ $M$ in the irreducible part of the two dimensional First Brillouin
Zone. In the range of energy that we have considered for this calculation, the
valence band is characterized by a lens at the top of the valence band at
$\Gamma$ that gets wider as the $C^{VI}$-cation gets heavier. The width of the
upper valence band (UVB) shrinks as the mass of the $B^{III}$ cation enhances.
For $B^{III}=In$ the small gap that separates the UVB from the intermediate
valence band (IVB) appears in the region around -5 eV. It is evident that the
UVB-width shrinks also with a bigger $C^{VI}$-anion mass as the optical gap
does in the whole series. In the conduction band (CB) energy region three
sub-bands are apparent separated by inner gaps. The gap around 7eV is of
particular interest as states appear within this energy region. We will
comment further on the behavior of these states below.

\section{The (001) ideal surface electronic band structure for the series of
chalcopyrites $CuB^{III}C^{VI}$ ($B=Al,Ga,In$; $C=S,Se,Te$).}

In Fig.1, we present the bands for $CuAlC_{2}^{VI}$, in Fig.2 for
$CuGaC_{2}^{VI}$ and in Fig. 3 for $CuInC_{2}^{VI}$. Electronic states appear
in the optical gap region and from 6-8 eV in the conduction band (CB). We
label the four states that we have found in the optical gap region as $E_{1}$
, $E_{2}$ , $E_{3}$ , $E_{4}$ and the three in the upper part of the CB as
$E_{5},$ $E_{6},$ and $E_{7}.$

\subsection{States within the optical gap energy-region.}

In the optical gap energy region of all the Cu-based chalcopyrites four states
occur. Two, $E_{1}$and $E_{2}$, are truly surface states and remain within the
gap energy region for the whole series. $E_{3}$ and $E_{4}$ behave
differently. Notice that the heavier the anion, the smaller the energy
difference between the top of the valence band and the surface states $E_{1}%
$and $E_{2}$. Nevertheless these two states remain well within the gap for the
whole series and almost at the middle of this energy interval. That is to say
that the gap also shrinks as the mass of the anion gets bigger for the same
cations in the perovskite. Also as the mass of the $B^{III}$ cation is
heavier, the gap decreases and the distance between the states $E_{1}$and
$E_{2}$ and the top of the valence band is shorter in energy. These features
are universal and can be appreciated quantitatively in Table I where we
present the energy of the four states within the optical gap at three
high-symmetry \ points in the two dimensional first Brillouin zone, namely,
$\Gamma$, $X$ and $M.$

\begin{center}

\begin{tabular}
[c]{|c|}\hline
element\\\hline
\\\hline
CuAlS$_{2}$\\\hline
CuAlSe$_{2}$\\\hline
CuAlTe$_{2}$\\\hline
CuGaS$_{2}$\\\hline
CuGaSe$_{2}$\\\hline
CuGaTe$_{2}$\\\hline
CuInS$_{2}$\\\hline
CuInSe$_{2}$\\\hline
CuInTe$_{2}$\\\hline
\end{tabular}%
\begin{tabular}
[c]{|c|c|c|c|}\hline
E1 [eV] & E2 [eV] & E3 [eV] & E4 [eV]\\\hline
$\Gamma$ \ \ \ \ \ \ X \ \ \ \ \ \ M & $\Gamma$ \ \ \ \ \ \ X \ \ \ \ \ \ M &
$\Gamma$ \ \ \ \ \ \ X \ \ \ \ \ \ M & $\Gamma$ \ \ \ \ \ \ X
\ \ \ \ \ \ M\\\hline
1.29 \ 1.26 \ 1.26 & 1.62 \ 1.41 \ 1.41 & 3.21 \ 3.21\ \ 3.21 & 3.30 \ 3.48
\ 3.48\\\hline
1.04 \ 0.99 \ 0.99 \  & 1.26 \ 0.99 \ 0.99 & 2.67 \ 2.74 \ 2.74 & 2.67 \ 2.74
\ 2.74\\\hline
0.63 \ 0.57 \ 0.57 & 0.84 \ 0.57 \ 0.57 & 2.01 \ 1.98 \ 1.98 & 2.01 \ 2.22
\ 2.22\\\hline
0.81 \ 0.81 \ 0.81 & 1.14 \ 0.87 \ 0.87 & 1.71 \ 1.74 \ 1.71 & 1.89 \ 1.98
\ 1.98\\\hline
0.69 \ 0.63 \ 0.57 & 0.81 \ 0.63 \ 0.57 & 1.44 \ 1.41 \ 1.41 & 1.44 \ 1.62
\ 1.62\\\hline
0.41 \ 0.35 \ 0.35 & 0.54 \ 0.35 \ 0.35 & 0.98 \ 0.92 \ 0.92 & 0.98 \ 1.13
\ 1.13\\\hline
0.57 \ 0.48 \ 0.48 & 0.75 \ 0.48 \ 0.48 & 1.62 \ 1.71 \ 1.71 & 1.74 \ 1.98
\ 1.98\\\hline
0.39 \ 0.18 \ 0.18 & 0.48 \ 0.33 \ 0.33 & 1.23 \ 1.50 \ 1.50 & 1.53 \ 1.68
\ 1.68\\\hline
0.30 \ 0.15 \ 0.15 & 0.42 \ 0.24 \ 0.24 & 1.18 \ 1.29 \ 1.29 & 1.32 \ 1.50
\ 1.50\\\hline
\end{tabular}

Table I- The energies at the high symmetry points $\Gamma,X,$ and $M$ for the
four states that appear in the optical gap region.

\end{center}

The states E$_{3}$ and E$_{4}$ lie very close to the bottom of the CB and
remain truly surface states for the whole series except for $CuInSe_{2}.$
There the two states become a resonance around $\Gamma.$ The surface state
E$_{4}$ enters into the CB at $X$ and $M$ for the three chalcopyrites
$CuAlC_{2}^{VI}$ studied here.

The composition of the surface states E$_{1}$ and E$_{2}$ is mainly of $s$ and
$p$ character. We also find a small contribution (less than $3\%)$ from states
of $d$-symmetry. This contribution decreases as the anion mass gets heavier
while at the same time the $p$-character of the state increases and its
$s$-character diminishes. The surface states E$_{3}$ and E$_{4}$ are more of
s-character but its $p$-contribution enhances as the anion mass gets bigger.
The $d$-contribution to this two states is very small or absent at all. We
quote the composition of these states for some cases in Table II.

\begin{center}
\bigskip%
\begin{tabular}
[c]{|c|}\hline
element\\\hline
\\\hline
CuAlS$_{2}$\\\hline
CuGaSe$_{2}$\\\hline
CuInTe$_{2}$\\\hline
\end{tabular}%
\begin{tabular}
[c]{|c|c|c|c|}\hline
E1 [\%] & E2 [\%] & E3 [\%] & E4 [\%]\\\hline
$s$ \ \ \ $p$\ \ \ \ $d$ & $s$ \ \ \ \ $p$ \ \ \ $d$ & $s$ \ \ \ \ $p$
\ \ \ $d$ & $s$ \ \ \ \ $p$ \ \ \ \ $d$\\\hline
43 \ 56 \ 1.7 & 47 \ 50 \ \ \ 2.8 & 67 \ 32 \ \ 0.0 & 78 \ 21 \ 0.2\\\hline
32 \ 67 \ 1.2 & 38 \ 59 \ 2.5 & 55 \ 45 \ 0.0 & 62 \ 37 \ 0.3\\\hline
22 \ 78 \ 0.7 & 28 \ 71 \ 1.0 & 55 \ 45 \ 0.0 & 55 \ 45 0.0\\\hline
\end{tabular}

Table II- Composition of some of the (100) surface states at $\Gamma$. The
percentages are rounded so that they do not exactly sum to 100\% in some cases.

\bigskip
\end{center}

Notice that as the mass of the cation ($B^{III}$) gets heavier a lens develops
around $\Gamma$at the bottom of the CB. Of special interest in this context is
$CuInSe_{2}$ since the two states $E_{3}$ and $E_{4}$ lie in the lens space.
For $CuInTe_{2}$ we find the $E_{4}$ state in the lens while the $E_{3}$ state
is not.

\subsection{States in the conduction-band-gaps energy region.}

Above the optical gap, within 10 eV from the top of the valence band (the
origin), there are two gaps in the CB. They do not shrink very much with the
enhancement of the anion mass conserving the cation ($B^{III}$) fixed but the
whole bulk band structure moves downwards to the origin so that the net effect
is that the optical gap diminishes as we already mentioned above. As a
consequence a third small gap appears around 10 eV. Notice also the
displacement of the structure towards lower energies as the anion gets heavier
for fixed cations. Three new surface states develop in this range of energy,
namely, $E_{5}$, $E_{6}$, and $E_{7}$. The energy at which these states occur
is smaller as the anion mass or as the cation $B^{III}$ gets bigger. We have
produced Table III to report on the energy at which these states appear.

\begin{center}%
\begin{tabular}
[c]{|c|}\hline
element\\\hline
\\\hline
CuAlS$_{2}$\\\hline
CuAlSe$_{2}$\\\hline
CuAlTe$_{2}$\\\hline
CuGaS$_{2}$\\\hline
CuGaSe$_{2}$\\\hline
CuGaTe$_{2}$\\\hline
CuInS$_{2}$\\\hline
CuInSe$_{2}$\\\hline
CuInTe$_{2}$\\\hline
\end{tabular}%
\begin{tabular}
[c]{|c|c|c|}\hline
E5 [eV] & E6 [eV] & E7 [eV]\\\hline
$\Gamma$ \ \ \ \ \ \ X \ \ \ \ \ \ M & $\Gamma$ \ \ \ \ \ \ X \ \ \ \ \ \ M &
$\Gamma$ \ \ \ \ \ \ X \ \ \ \ \ \ M\\\hline
7.05 \ 6.90 \ 6.90 & 7.05 \ 6.99 \ 6.99 & 8.28 \ 8.34 \ 8.34\\\hline
6.39 \ 6.40 \ 6.40 & 6.52 \ 6.52 \ 6.52 & 7.64 \ 7.66 \ 7.66\\\hline
5.52 \ 5.52 \ 5.52 & 5.64 \ 5.64 \ 5.64 & 6.72 \ 6.69 \ 6.72\\\hline
6.09 \ 5.91 \ 5.91 & 6.09 \ 6.06 \ 6.06 & 7.38 \ 7.38 \ 7.38\\\hline
5.55 \ 5.55 \ 5.55 & 5.70 \ 5.70 \ 5.70 & 6.87 \ 6.87 \ 6.87\\\hline
4.89 \ 4.91 \ 4.91 & 5.02 \ 5.06 \ 5.06 & 6.10 \ 6.11 \ 6.11\\\hline
4.32 \ 3.49 \ 3.49 & 4.86 \ 4.53 \ 4.53 & 6.72 \ 6.66 \ 6.66\\\hline
5.28 \ 5.07 \ 5.07 & 5.28 \ 5.16 \ 5.16 & 6.30 \ 6.27 \ 6.27\\\hline
4.80 \ 4.80 \ 4.80 & 4.94 \ 4.92 \ 4.92 & 5.91 \ 5.91 \ 5.91\\\hline
\end{tabular}

Table III- The same as Table I for the three states that appear in gap region
in the conduction band.

\end{center}

In the series $CuAlC^{VI}$ the state $E_{5}$ lies in the CB for the whole
interval $X-\Gamma-M$ for $C^{VI}=S$ but emerges from the bulk band energies
for $C^{VI}=Se$ and is neatly detached from the CB for $C^{VI}=Te.$ $E_{6}$ is
a resonance around $\Gamma$ but emerges as a pure surface state around $X$ and
$M$ for the first element of this sub-series but is a pure surface state for
the other two. $E_{7}$ remains as a pure surface state in the whole branch of
the 2D-FBZ reported here for the three elements.

In the series $CuGaC^{VI}$ the behavior of the three surface states $E_{5}$,
$E_{6}$, and $E_{7}$ is approximately the same as for the $CuAlC^{VI}$one.

The series $CuInC^{VI}$ presents a different behavior. The state $E_{5}$ lies
in the upper border of a lens that develops in the second inner gap in the CB
and it is therefore to be consider as a resonance for $C^{VI}=S$ . For
$C^{VI}=Se$, $E_{5}$ gets through the gap towards higher energies and enters
the CB. But for $C^{VI}=Te$, $E_{6}$ is a pure surface state for the whole
interval $X-\Gamma-M$.

Also the state $E_{6}$ behaves specially. In $X-\Gamma-M$, it is a resonance
in almost the whole interval (see Fig. 3 for details) for $C^{VI}=S,Se$. But
for $C^{VI}=Te$ it emerges as a surface state in the whole interval presented
here. $E_{7}$ remains essentially as a pure surface state in $X-\Gamma-M$ for
the whole series. The three $E_{5}$, $E_{6}$, and $E_{7}$ states are of almost
exclusively of p-character.

Finally, two more surface states emerge in the energy region under study (less
than 10 eV) since as a consequence of the displacement towards lower energies
of the bulk bands (shadowed areas in the figures 1-3) as the mass of the
$B^{III}$ cation and of the anion gets heavier, a new third gap appears in the
CB below 10 eV. This is apparent in $CuAlTe_{2}$ in Fig.1. $Te$ is the heavier
anion of the sub-series $CuAl.$ In the second series $CuGa$ in Fig.2, the
third gap in the CB is seen for the anion $Se$ at around 10 eV and appears
around 9 eV for the anion $Te$ and moves towards 8 eV for $CuInTe_{2}$. The
energy at $\Gamma$ for some of these states is presented in the next Table IV.
The composition at $\Gamma$of both states that appear at this energy is almost
exclusively of p-orbitals, with a very minor contribution from s-orbitals and
a tiny one from d-orbitals.

\begin{center}%
\begin{tabular}
[c]{ccc}
& E$_{8}$ & E$_{9}$\\
CuInS$_{2}$ & 9.48 & 9.83\\
CuInSe$_{2}$ & 9.12 & 9.48\\
CuInTe$_{2}$ & 8.34 & 8.78
\end{tabular}

Table IV. The surface states at higher energies in the conduction bands
\end{center}

Just for completeness, we mention that some surface state appear in the range
of energy studied here in some of the lenses in the valence band region that
are apparent from Figs. 1-3.

\section{The (001) ideal surface local density of states for the series of
chalcopyrites $CuB^{III}C^{VI}$ ($B=Al,Ga,In$; $C=S,Se,Te$).}

In Fig. 4 we present the (001) surface local density of states (LDOS) for the
series of chalcopyrites studied here. The points indicate the optical gap. The
peaks in the optical gap indicate the spectral weight of the surface states in
the optical gap energy region, namely $E_{1}$ , $E_{2}$ , $E_{3}$ , and
$E_{4}.$

The CB present two minima associated with the two inner gaps that we have
presented above. Also its evolution with the anion and $B^{III}$ cation masses
is apparent. The shrinking of the optical gap as the anion mass, on one hand,
and the cation one, on the other, increases is clearly seen in this figure as
well.\textbf{ }Between -6 and -8 eV the LDOS has the highest spectral weight,
a common feature to all the nine Cu-based chalcopyrites due to the
contribution of $d-Cu$ orbitals. Notice that this peak becomes thinner and
higher as the mass of the $B^{III}$ cation enhances. The LDOS at this energy
presents two peaks for $C^{VI}=S,Se$ and only one for $C^{VI}=Te.$

\section{Fisim states}

Non dispersive states have been experimentally found in (100) CdTe
\cite{fisimniels}. These states were described as bulk states induced by the
presence of a surface \cite{daniel1y2}. It was found that these states are
more general and predicted to exist at interfaces as well
\cite{fisiminterfaces}. The states were consequently suggested to be named
Frontier Induced Semi-Infinite Medium (FISIM) states \cite{european}. In these
papers, the FISIM states are shown to be built from spectral weight that
accumulates at a certain fixed energy from states with the same crystal
momentum $k$ and therefore their occurrence does not violate the conservation
of the number of states when integrating in the energy interval of the whole
valence band at constant $k$. A detailed study of the valence-band electronic
structure for ZnSe(001) is presented in reference \cite{umkklapp}. The
non-dispersive states are found. It seems that this work is consistent with
these states being bulk states (existing not only in the surface region) but
induced by the particular surface in the direction studied, that is to say
FISIM states. Nevertheless whether their detailed interpretation does violate
or not the conservation theorem quoted above has not been studied in detail.
We will examine further this point somewhere else. In the energy interval
studied here we did not find FISIM states for the (001) direction as it was
found in the (112) surface electronic band structure for $CuInSe_{2}$
\cite{european}.

\section{Conclusions}

We have used the SGFM method and a tight-binding hamiltonian to calculate the
ideal (001)-surface electronic band structure and the Local Density of States
(LDOS) for the series of Cu-based chalcopyrites $CuB^{III}C^{VI}$
($B=Al,Ga,In$; $C=S,Se,Te$). The general characteristics are similar although
important differences occur in the bands as in the LDOS. Within the optical
gap energy region we find four surface states of mainly s-p character and in
the CB region we find three surface states of predominantly p-character. The
bulk projected electronic bands evolve in the series as the mass of the anion,
on one side, and the one of the $B^{III}$ cation on the other, gets heavier,
moving the conduction band (CB) as a whole towards lower energies. We found
two consequences of this fact. First, the optical gap shrinks and second a
third inner gap appears in the higher energy region of the CB. Two new surface
states of predominantly p-character exists in this new gap for the
chalcopyrites that do show this third inner gap within 10 eV from the origin.
Presumably the gap is a common fact to the whole series but lies at higher
energies in the rest of the members of the series. Finally, we did not find
FISIM states for the (001) surface band structure as it was found for the
(112) one previously for $CuInSe_{2}$ \cite{european}.

\bigskip

\bigskip

\bigskip

FIGURE CAPTIONS

FIG.1 The electronic band structure for the sub-series $CuAlC^{VI}.$ The
surface states are indicated as dotted lines.

FIG.2 \ We present in this figure the electronic band structure for the
sub-series $CuGaC^{VI}.$ The conventions are the same as in Fig. 1.

\bigskip FIG.3 The electronic band structure for the sub-series $CuInC^{VI}.$

FIG.4 The Local Density of States (LDOS) at the (001) surface layer for the
whole series of Cu-based chalcopyriyes.

\end{document}